# EXTENSION OF THE EQUIVALENCE BETWEEN ACCELERATION AND GRAVITATAION

Erez M. Yahalomi

## I. INTRODUCTION

The theory of special relativity [1] deal with the differences in physical element like time, mass, energy and length on systems that move in uniform velocities relative to each other. Conventional special relativity argues there is no preferred frame. And the measured magnitude of the physical elements are relative to the observed frame.

In the theory of general relativity there is a privileged inertial frame the magnitude of the physical elements depends on the gravity and mass surround them .For example the time change of an object in general relativity can be calculated absolutely from the mass and gravitation influence on the object.

Email:erezsu@hotmail.com

There were several attempts to show that there is a privileged frame [2-7].

We found a new definite explanation and cause , for a distinction of the privileged inertial frame between two frames, which move in uniform velocity relative to each other. Applying this physical explanation to the phenomena treated in special relativity equations result in modified equations that gives in some cases new results and can explain experiment results that are not matched with calculations of conventional special relativity equations.

The physical difference we found between two frames which move in uniform velocity relative to each other is that only one of the frame experienced acceleration which brings the frame or object to uniform velocity relative to the other .The magnitude of the relativistic phenomena depend on the length of time the object accelerated in the past .The period of acceleration in the past can looked as phase transition which cause the special relativity phenomena in the frame or object that move later in uniform velocity. **This can be regarded as an extension of the general theory of relativity equivalence principle, which state that acceleration equal gravitation. Like the gravity in the theory of general relativity which influenced time dilation, an acceleration period in the past which bring frame to uniform velocity cause phenomena like time dilation,** A frame which did not experience acceleration period in the past to reach the uniform velocity may look like moving when observed from other frame but does not have relativistic effects.

We analyze Keating and Haflee experiment [8-9] of flying atomic clock around the world and comparing the time difference between the clock which stayed at the surface and the clocks which circled earth , and find the experiment results are in agreement with our theory.

We give here modified equations for space time coordinates ,additions of velocities, mass and energy.

Finding that the acceleration period in the past cause the special relativity effect in uniform velocity moving, frame bring us to the conclusion that even an object which looked at rest in one scale can be considered to move, on a bigger scale.For example a star looked to be at rest for an object on it's surface but it moves on a scale of a galaxy. This means that in the past the star experienced acceleration which bring it to this velocity. This leads us to the new findings of a Universal Inertial Frame, which is the absolute rest frame for all objects. In the Big bang theory this frame can be considered as the Big bang location before the expansion.   We suggest a way to find the earth velocity  relative to the Universal Inertial Frame.

We start with a thought experiment, Consider two π mesons one is at high altitude, and the second meson is on earth surface. Assume the first meson moves towards earth at a constant velocity v. If its proper lifetime is $\tau_1$ its lifetime measured from the surface considers as the inertial frame is $\tau_1'$.

[1]

$$\tau_1' = \frac{\tau_1}{\sqrt{1-\frac{v^2}{c^2}}}$$

$\tau_1$ is also the lifetime in the laboratory frame, of the π meson which stayed on earth surface. This imply the possibility that on a very close distance to earth the meson form the high altitude after traveling to this point would still exist where the surface meson would have already decayed. According to special relativity we can assume the meson at high altitude is at rest and the meson on earth surface travel towards the first meson at velocity -v. In this case the proper lifetime of
the meson on the surface is defined as $\tau_1 = \tau_2$.
Considering the high altitude meson reference frame as the laboratory reference frame. The lifetime of the surfaced meson is measured as:

[2]

$$\tau_2' = \frac{\tau_2}{\sqrt{1-\frac{v^2}{c^2}}}$$

Where $\tau_2'$ is the earth surface meson lifetime in the, high altitude meson laboratory frame. The lifetime of the surface meson is considered to be longer. Then when the distance between the two mesons will get close the meson from the high altitude would decay and the surface meson would still exist.
We received two different results describing the same physical state.

Our solution to the problem comes from the discovery that there is a difference between the two frames. Only one of the frames accelerated to reach velocity V. The acceleration period is the source to the special relativistic phenomena.

The following term is presented:

[3]

$$V = \int_0^{t_1} a(t_p)\,dt_p$$

a(tp) is the time dependent acceleration, tp present time in the past, t1 is the time until the body reaches velocity V and stops accelerating. The variable tp is introduced to emphasize the acceleration in the integral of eq.3 is in the past, instead of tp we can substitute the variable t as long as we remember that the boundaries of the integral are in the past before the measurement of the special relativity phenomena. The integration in eq.3, between 0 to t1 is between two events in the past. for example the acceleration accrued between 3/2/2001 to 4/4/2001. Later the object or frame stopped accelerating and continue to move in uniform velocity till the present. The measurements of the special relativity phenomena are when the object move in uniform velocity.

[4]

$$x' = \frac{x - \left(\int_0^{t_1} a(t_p)\,dt_p\right) t}{\sqrt{1 - \left(\dfrac{\int_0^{t_1} a(t_p)\,dt_p}{C}\right)^2}}$$

$$y = y$$
$$z = z$$

[5]

$$t' = \frac{t - \left(\int_0^{t_1} a(t_p)\,dt_p\right)x/C^2}{\sqrt{1 - \left(\dfrac{\int_0^{t_1} a(t_p)\,dt_p}{C}\right)^2}}$$

x,t are the coordinates in the inertial frame. x',t' are the coordinates in a moving reference frame that moves in uniform velocity V relative to the inertial frame.The coordinates x,t,x',t' are in the present .The terms a(tp), dtp inside the integral are in the past.t1 is the time until the object reach uniform velocity.

The coordinates equations where x',t' are considered at rest are identical to Eq. [4,5] because the inertial frame is always the frame which has not accelerated and the frame which has relativistic effects is the frame which accelerated to velocity V.

We can use our result to explain the famous twin paradox.The twin paradox considers two twin, one is flying into space and returns to earth, the other stays on earth. The one who flew into space aged less. The usual explanation [10] is that the twin who flew into space experienced acceleration so his time changed slower.From our result we can conclude that the time difference between the two twins is not relay solely on the act of acceleration which can take an infinitesimal time and it's influence would be infinitesimal . The time period ofacceleration is important too because the speed the twin reaches determine the factor of the time difference. The time difference between the two twins is depends on this factor multiply by the amount of time the flying twin moves in uniform velocity. The faster is the uniform velocity, reached by this acceleration and the longer is the time this twin move in uniform velocity, the larger is the time difference between the two twins when they meet again, as seen from the second equation in eq .5.

**The equation for the total energy**

The modified equation for total energy is depend on the acceleration period in the past. E0 is the total energy in the inertial frame.

$$E_0 = m_0 c^2$$

$$E = \frac{E_0}{\sqrt{1 - \left(\frac{\int_0^{t_1} a(t_p) dt_p}{C}\right)^2}}$$

[6]

The equation for kinetic energy is:

[7]

$$E_k := \frac{m_0 c^2}{\sqrt{1 - \left(\frac{\int_0^{t_1} a(t_p) dt_p}{C}\right)^2}} - m_0 c^2$$

**The equations of the addition of velocities in one dimension.**

Let S' frame moves with velocity V relative to S and an object A move with Velocity u' relative to S', object A accelerated from rest in S' frame at t2 and stopped accelerating at t3.

[8]

$$u = \frac{\int_{t2}^{t3} a(t_p) dt_p + \int_0^{t_1} a(t_p) dt_p}{1 + \int_{t2}^{t3} a(t_p) dt_p \cdot \int_0^{t_1} a(t_p) dt_p / C^2}$$

Where

$$u' = \int_{t_2}^{t_3} a(t_p) dt_p \quad V = \int_{0}^{t_1} a(t_p) dt_p$$

u is the velocity of object A in the inertial frame s. Velocity V is equal to the integration of the acceleration period in the past from time 0 to time t1. u' is equal to the acceleration period in the past from time t2 to time t3.

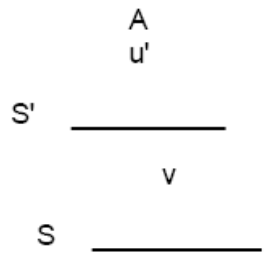

```
            A
            u'
    S'  _________

                V
    S   _________
```

Fig 2: frame S' moves with velocity V relative to the inertial frame S. Object A moves with velocity u' relative to S'.

**We present the equation for mass transformations :**

Let $m_0$ be a rest mass and m is the mass which accelerated to velocity V.

We obtain.

$$m = \frac{m_0}{\sqrt{1 - \left(\frac{\int_{0}^{t_1} a(t_p) dt_p}{C}\right)^2}} \qquad [9]$$

The change in the mass magnitude of a uniformly moving mass compared to a mass at rest depends in the period of acceleration the moving mass experienced in the past.

Let $m_1$ be a mass at rest in reference s' which moves with velocity V, compared to Inertial frame s. Assume mass $m_1$ accelerated to velocity u' relative to frame s' There are two options for moving with velocity u' relative to s', one moving with Velocity u', the other with velocity -u'.

In the conventional special relativity equations for both cases we have

$$m_2 = \frac{m_1}{\sqrt{1 - \frac{|u'|^2}{C^2}}}$$

$m_2$ is the mass moving with velocity $|u'|$ relative to s'. But according to our postulate that the relativistic effects depend on the acceleration that brings the object to velocity V, we must consider the velocity compared to the inertial frame s. We use the addition of velocities equation (eq.7)

For velocity equals u'

$$u = \frac{\int_{t2}^{t3} a(t_p)dt_p + \int_0^{t_1} a(t_p)dt_p}{1 + \int_{t2}^{t3} a(t_p)dt_p \cdot \int_0^{t_1} a(t_p)dt_p / C^2} \quad (i) \quad\quad [10]$$

u is the velocity compared to the inertial frame s,

For velocity equals -u'

$$u = \frac{-\int_{t2}^{t3} a(t)dt + \int_0^{t_1} a(t)dt}{1 - \int_{t2}^{t3} a(t)dt \cdot \int_0^{t_1} a(t)dt / c^2} \quad (ii)$$

Mass $m_2$ is mass $m_0$ which accelerated to uniform velocity u relative to frame s

[11]

$$m_2 = \frac{m_0}{\sqrt{1-\frac{u^2}{C^2}}} \qquad m_1 = \frac{m_0}{\sqrt{1-\left(\frac{\int_0^{t_1} a(t_p)dt_p}{C}\right)^2}}$$

$$m_2 = \frac{m_1 \sqrt{1-\left(\frac{\int_0^{t_1} a(t_p)dt_p}{C}\right)^2}}{\sqrt{1-\frac{u^2}{C^2}}}$$

Since $m_2$ is dependent on the acceleration from 0 velocity, it has different values When observing from s' frame if the velocity compares to frame s' is u' or -u'. Assume mass m is accelerated to reach velocity u' relative to earth , first in the direction of earth's rotation , and in the second case to velocity u' relative to earth in the opposite direction of earth rotation. The earth rotation velocity V ,is assumed in this example as the velocity relative to the inertial frame, m0 is the rest mass at the inertial frame . m1 is the rest mass on the surface of the earth and m2
will be different in the two cases.

For u'=V

$$u = \frac{2V}{1+\frac{V^2}{C^2}}$$

[12]

And m2 > m1  >  m0

For u'= -V:

u=0  and m2 = m0 <m1 .

We obtained the result that there are cases where a moving mass is smaller than a rest mass on earth.

We obtained that special relativistic phenomena derived from a time region where the object accelerated to reach its uniform velocity. We want to find the absolute inertial frame, which didn't experience any acceleration .This absolute frame is an inertial frame to the all universe (on the special relativity aspect of inertial frame) .

We define the Universal Inertial Frame as the frame which all the objects that compared to this frame are considering as moving, experienced acceleration in the past.

We can conduct an experiment where we check mass in continuity of velocities and in different directions, to find the universal inertial frame velocity. If we move the measured mass in a velocity which is opposite to it's velocity relative to the universal inertial frame both velocities would cancel and we get the mass magnitude in the universal inertial frame, this is the minimum mass value obtained in a measurement .The magnitude of this opposite velocity is equal to the velocity of the system relative to the universal inertial frame.

Another method to obtain the velocity of the universal inertial frame is by Measuring the rate of an atomic clock at different speeds and in different directions this gives different clock rates. When the rate is the highest, after eliminating the effects of the earth non-inertial frame, we get the universal inertial frame velocity.

$$dt = \frac{dt_0}{\sqrt{1-\left(\frac{\int_0^{t_1} a(t_p)dt_p}{C}\right)^2}} \qquad [13]$$

$$u = \int_0^{t_1} a(t_p)dt_p$$

u is the velocity relative to the universal inertial frame,
dt is the rate of the moving clock, dt0 is the rate of the clock at rest in the universal inertial frame .For v>u' while v is the velocity of earth relative to the universal inertial frame dt will be higher than the clock rate at earth. Though in some cases a clock which move relative to earth surface its time rate can be measured on the surface to be higher than a clock at rest on the surface. The best place to find the universal inertial frame is in space where the influence of the rotational movement of earth is reduced.

The Universal inertial frame can be considered as the Big Bang source before the expansion. To those, which are not convinced in the theory of the Big Bang, a future experiment for measuring the Universal inertial frame with matter from several locations in the Universe could determine if the Universe really expanded from one location. If the Universe expands from one location, rate measurement of an atomic clock made of matter from one place in the Universe when compared to the rate of an atomic clock made of matter from another place in the Universe, would show that the highest rate of both clocks when measured at the same place is obtained at the same direction and velocity, then both matters come from the same origin. If the result shows two different directions or velocities, then there are two origins.

### II. A suggested practical experiment by mosbauer effect

Suggested practical experiment using Mossbauer effect. Because of the small natural width of high frequency gamma radiation Mossbauer effect is an excellent tool to verify relativistic effects. The following experiment is suggested.

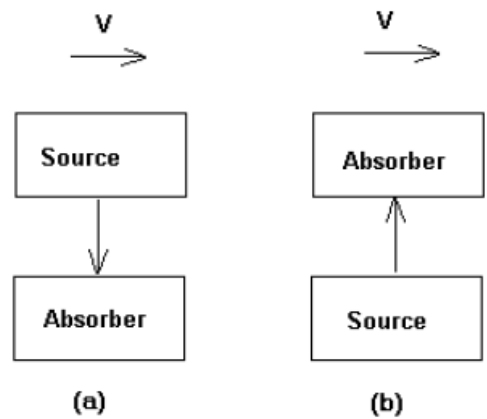

Fig. 2. Two sets up for measurement of γ ray frequency shift due to relativistic effects by Mossbauer experiment.
(a) A source of gamma rays moves in a straight line in a constant velocity. The absorber absorbs γ rays emitted perpendicular to the source movement.
(b) An absorber of γ rays moves in a straight line. The absorber absorbs γ rays emitted perpendicular from stationary source.

On the first measurement, the source of gamma rays moves in a straight line. The absorber absorbs the gamma rays emitted perpendicular to the source movements. On the second measurement the absorber of gamma rays moves in a straight line. The absorber absorbs the gamma rays emitted perpendicular from the stationary source. the usual special relativity predicted to measure a frequency red shift, due to transverse Doppler effect in both cases.

[14]

$$\Delta v = \frac{V^2}{2c^2} v$$

Where $$V = \int_0^{t_1} a(t_p) dt_p$$

Δv is the frequency red shift, V is the velocity between the source and the absorber, V is is the gamma ray frequency, C is the speed of light.
According to our theory, in the first case the measured frequency is the same velocity V. In the second case, our theory predicted a blue shift. In this case the absorber accelerated before reaching velocity V, which according to Eq. (5) causes a time dilation, which is interpreted as a red shift in the natural width of the absorber. The source was not accelerated though according to Eq. (5) there is no time dilation or frequency shift in its natural width. Therefore, gamma rays from the source measured at the absorber, will appear to be blue shifted. This phenomenon has not been noticed before, since in the former experiments [11], the source always moved and the absorber was stationary.

### III. Experiment results supported the universal inertial frame

The reason this phenomena is not widely known is ,there are rare experiments that the relativistic effects are measured from the objects (particles) that accelerated to the uniform velocity close to the speed of light. Usually the special relativity phenomena are measured from the laboratory frame which is close to the inertial frame.Where conventional special relativity equations and our equations give close results. An experiment that measures the relativistic effects from the frame of the object, which accelerated to uniform velocity, is Hafele and Kiting experiment. Hafele and Kiting experiment [9] strongly support the existence of a Universal Inertial frame. In this experiment four cesium beam atomic clocks were flown around the world twice, once eastward and once westward to test the theory of relativity. The time difference between the flying clock and the clock, which stayed on the earth's surface, was compared. The flying clock lost 59 - + 10 nanosecond during the eastward direction and gained 273 - + 7 nanoseconds during the westward trip. If we analyze a airplane flight in around the world trip it is seen that most of the time the airplane fly in uniform velocity .In order to fly in circular, track the airplane seldom make an angular correction of it's flying direction, this and the time the airplane change it's height because of elevation forces are the only time the airplane experience radial acceleration, which is negligible compare to the flight period.
The result can only be explained by taking into account the universal inertial frame. The earth has a rotational velocity, in the eastward direction flight the velocity of the plane is positively added to the earth velocity and increased the velocity of the clock on the airplane relative to the universal inertial frame and the clock lose time relative to the clock which stayed on the surface .In the westward trip the airplane velocity added negatively to the earth velocity this decrease the velocity of the clocks on plane relative to the universal inertial frame and the clocks gained time relative to the clock which stayed on the surface .

In the conventional special relativity the time difference is calculated as taking the clock on the surface as being in an inertial frame, the frame can be considered as inertial in infinitesimal increment. The airplane is considered to move in velocity v in the eastward trip and in velocity v but in the opposite direction in the westward trip.The time dilation in conventional special relativity depend on the square of v, eq .1. In conventional special relativity in both directions the clocks on the plane lose time relative to the clock on the surface, this is not in agreement with the experimental results.

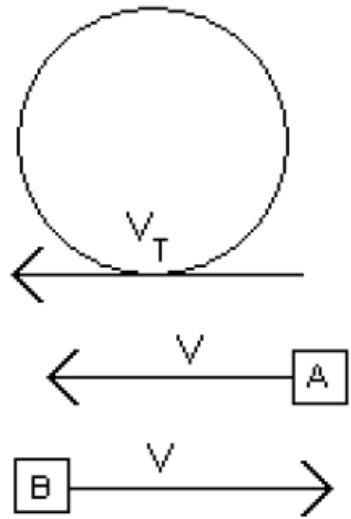

Fig 3: The circle represent earth with tangential velocity $V_T$. A is a airplane flying in Velocity V relative to earth's surface. B is the flying airplane that is flying in velocity V Relative to earth's surface but in the opposite direction to A.

Considering the North pole as inertial frame [8] give a closer result to the experiment, although this frame does not gives a direct comparison between the clock in the surface frame and the flying clock like in the twin paradox. There are other difficulties; the experiment showed no symmetry in the mean errors between the experience and the predicted results at the eastward and westward directions. For the westward direction the mean difference between the predicted and the experiment results is 0.7% and the mean difference in the eastward direction is 15%. This implies the inertial frame considered is not the exact inertial frame. Although Hafele and Keating predicted calculations gave a good approximation to the experience result by using the north pole as the inertial frame in the calculations and not only conventional special relativity, Calculations of the exact inertial frame should take into account the Universal Inertial Frame.

The notation tp was introduced only to emphasize that the time in the relevant interaction is in the past.**It is possible to write the same equations with the variable t as long as we remember that the integration limits on a(t) are in the past ,before the measurements when the object moves in uniform velocity.**It is also possible to replace the integration term $\int_0^{t_1} a(t_p) dt_p$ by the velocity variable V.As long as we identified between the cases where the integration is replaced by V and the cases where the integration is replaced by zero, this is the cases when a(tp) is equals zero.The substitution of V is an approximation because the measured V in the laboratory inertial frame is a close approximation to the measured velocity from the Universal inertial frame.

In summary, the relativistic effects of special relativity are derived from integration of the acceleration over a period of time before the object move in uniform velocity and the measurement of the special relativity phenomena begin.  This complements full explanation of  the twin paradox.   A universal reference frame is obtained. It implies that on a proper direction there is a range of velocity in which the rate of time on a moving frame can increase, the magnitude of mass can decrease. While in the standard special relativity the time rate of a moving frame can only decrease compared to rate on earth and the magnitude of mass can only increase . An experiment based on the rate's measurement of atomic clock as we described could discover the speed the Earth moves relative to the Universe before the Big Bang.

**Acknowledgment.** The author likes to thanks G.Iosilevskii for useful discussions on flight mechanics.